\documentclass[journal, final]{IEEEtran}
\usepackage{cite}

\ifCLASSINFOpdf
\else
\fi
\usepackage{graphicx}
\usepackage{amssymb}
\usepackage{multirow}
\usepackage[cmex10]{amsmath}
\usepackage{algorithmic}

\begin{document}
%
\title{Performance of a Concurrent Link SDMA MAC under Practical PHY
Operating Conditions}
%
%
%

\author{Pengkai Zhao, Babak Daneshrad, Ajit Warrier, Weijun Zhu,
        Oscar Takeshita 
 \thanks{Copyright (c) 2010 IEEE. Personal use of this material is permitted.
        However, permission to use this material for any other purposes must be obtained from the
        IEEE by sending a request to pubs-permissions@ieee.org.}
  \thanks{Pengkai Zhao and Babak Daneshrad are with the Department
 of Electrical Engineering, UCLA, Los Angeles, CA.}
 \thanks{Ajit Warrier, Weijun Zhu and Oscar Takeshita are with Silvus Technologies, Inc., Los Angeles, CA.}}
\maketitle

\begin{abstract}
Space Division Multiple Access (SDMA) based Medium Access Control (MAC)
protocols have been proposed to enable concurrent communications and improve
link throughput in Multi-Input Multi-Output (MIMO) Ad Hoc networks. For the
most part, the works appearing in the literature make idealized and
simplifying assumptions about the underlying physical layer as well as some
aspects of the link adaptation protocol. The result is that the performance
predicted by such works may not necessarily be a good predictor of actual
performance in a fully deployed system. In this paper we look to introduce
elements into the SDMA MAC concept that would allow us to better predict
their performance under realistic operating conditions. Using a generic
SDMA-MAC we look at how the network sum throughput changes with the
introduction of the following: $(a)$ use of the more practical MMSE algorithm
instead of the zero-forcing or SVD based nulling algorithms used for receive
beamnulling; $(b)$ impact of channel estimation errors; $(c)$ introduction of link
adaptation mechanism specifically designed for concurrent SDMA MACs; $(d)$
incorporation of TX beamforming along with RX beamnulling. Following on the
transmission window during which concurrent transmissions are allowed by the
MAC, we qualify the impact of each of these four elements in isolation. At
the conclusion, the performance of a system that incorporates elements $a-d$ is
presented and compared against the baseline system, showing an improvement
of up to 5x in the overall network sum throughput.
\end{abstract}


%
\IEEEpeerreviewmaketitle

\section{Introduction}

Networks of MIMO (Multi-Input Multi-Output) enabled nodes can use advanced eigen-beamforming and
beamnulling techniques to enable concurrent communications and increase
overall network throughput. This technique is loosely
referred to as space division multiple access and several medium
access control (MAC) protocols have appeared in the literature that can
deliver concurrent transmissions in an Ad Hoc network of multi-antenna, MIMO,
enabled nodes \cite{1_SPACEMAC, 2_MIMOMAN, 3_NULLHOC, 4_nullhoc, 5_MIMA_MAC}.

Although SDMA (space division multiple access) and concurrent links have been well studied in cellular
networks (see \cite{VT_SDMA} and the references therein), it is still a challenging
problem in Ad Hoc networks. Initially, SDMA and concurrent links were
utilized in Ad Hoc networks via a simple abstract model called Degree of
Freedom (DOF) \cite{6_Wireless_Comm_David, COOP_MAC}. This model uses the number
of antennas to represent the number of concurrent links in the network. It
assumes that the concurrent links are perfectly separated and do not
interfere with one another. As such, the DOF model ignores all physical
layer (PHY) impairments. At the same time, using TX/RX beamforming, the
SPACEMAC, MIMAMAC and NullHoc protocols \cite{1_SPACEMAC, 2_MIMOMAN, 3_NULLHOC, 4_nullhoc, 5_MIMA_MAC} have been proposed to
support concurrent links in Ad Hoc networks. These protocols assume that the
first node to ``win'' the contention window will use an omni directional
radiation pattern, but other, secondary, users will use TX beamforming to
ensure that newly accessing link will not introduce any interference to
existing links. As a result, the throughput of existing links is not
affected, and additional network throughput can be had as a result of the
newly formed concurrent links. Although this idea works well under an ideal
MIMO system model, its performance is significantly affected by physical
layer constraints and imperfections (e.g., channel estimation error, absence
of link adaptation, etc.). In this paper we focus on the concurrent
transmission window within a generic SDMA-MAC that is similar in its
construction to SPACEMAC and NullHoc. These MACs were proposed for Ad Hoc
networks and typically use signaling during the contention window to
determine the TX {\&} RX beam patterns to be used during the concurrent
transmission windows. They typically make idealized and simplified
assumptions that will impact their performance during the concurrent
transmission window. In this paper our aim is to migrate an idealized
SDMA-MAC system, such as the ones found in \cite{1_SPACEMAC, 2_MIMOMAN, 3_NULLHOC, 4_nullhoc}, towards a more realistic
one that incorporates (a) channel estimation error, (b) the use of a more
practical MMSE detection algorithm, (c) incorporation of link adaptation,
and (d) combined TX and RX beamforming techniques. Our study uses the
generic SDMA-MAC protocol presented in section II as a baseline, but the
results could be easily extended to other MACs with a similar structure. For
each of the elements (a) through (d) we compare the performance of the
baseline SDMA-MAC during the concurrent transmission window with and without
the proposed modification. We then combine all the changes together and
compare the performance of the resulting ``practical'' MAC with the baseline
system. To this end, the paper will be organized as follows. Section II
introduces our system model, the baseline SDMA-MAC, and the simulation setup.
Section III describes each of the four elements of our proposed
modifications and the associated performance gain/loss for each element in
isolation. In section IV we provide side by side comparisons of the baseline
concurrent SDMA-MAC, the realistic variant of the concurrent SDMA-MAC which
includes elements a-d, and a non-concurrent MAC that utilizes MIMO links.
The paper is then concluded in Section V.

\section{System Description}

This paper focuses on a single hop Ad Hoc network, where each node is within
the transmission range of all other nodes. There are a total of $K$ concurrent links simultaneously transmitting in the network,
labeled as link $L_1$ to link $L_K$. The TX node and RX node involved in
link $L_q$ are denoted as $T_q$ and $R_q$, respectively. Every node is equipped with
$N_A$ antennas, and all packets are modulated
using OFDM (Orthogonal Frequency Division Multiplexing), where the number of subcarriers is $N_C$. The TX power per
node is the same and is denoted by $P_T$.  The fast fading
channel from the TX node $T_q$ to the RX node $R_q$ at the $i$th
subcarrier is ${\rm {\bf H}}_{R_q ,T_q } (i)$, which is an $N_A \times N_A $
matrix of complex Gaussian random variables with zero mean and unit
variance. $G_{R_q ,T_q } $ is
the path loss from node $T_q $ to node $R_q $. For simplicity, we assume that each link uses a single
spatial stream. However, our discussions can be easily generalized to other
cases where some links might use multiple spatial streams. We denote the
power normalized $N_A \times 1$ TX vector at the $i$th subcarrier of node
$T_q $ as ${\rm {\bf W}}_{T_q } (i)$ (power normalized implies that ${\rm
{\bf W}}_{T_q }^H (i){\rm {\bf W}}_{T_q } (i) = 1)$. Similarly, the RX
vector at the $i$th subcarrier of node $R_q $ is ${\rm {\bf W}}_{R_q } (i)$
subject to ${\rm {\bf W}}_{R_q }^H (i){\rm {\bf W}}_{R_q } (i) = 1$. Also, the transmitted QAM symbol at each subcarrier is zero mean and unit variance. Finally, we use ${\rm {\bf
A}}(i)$ to represent the matrix corresponding to the $i$th subcarrier, and
${\rm {\bf A}}(i,j)$ is the $j$th column of matrix ${\rm {\bf A}}(i)$. $[
\cdot ]^H$ and $[ \cdot ]^T$ are Hermitian and transpose calculation.

\subsection{Overview of the Generic SDMA-MAC}
Our baseline SDMA-MAC is designed to represent a class of MACs such as
SPACEMAC \cite{1_SPACEMAC, 2_MIMOMAN} and NullHoc \cite{3_NULLHOC, 4_nullhoc}, it is built on the principal that links
have an access hierarchy in that the newly accessing link should cause no interference to the existing links. This
process is described mathematically in the following subsection.

\subsection{Mathematical Description of a Generic SDMA-MAC}
Let the access order of $K$ concurrent links in the network range from $L_1
$ (link 1) to $L_K $ (link $K$). Currently the first $(Q - 1)$ links have
accessed the channel, and now we look at the access process of link $L_Q $,
$1 \le Q \le K$. According to our generic SDMA-MAC protocol, during the
concurrent transmission window, link $L_Q $ should use a TX vector, ${\rm
{\bf W}}_{T_Q } (i)$, that is orthogonal to the existing links' RX vectors
${\rm {\bf W}}_{R_q } (i)$, or equivalently:
\begin{eqnarray}
\label{eq1}
{\sqrt {P_T
G_{R_q ,T_Q } / N_C } {\rm {\bf W}}_{R_q }^H (i){\rm {\bf H}}_{R_q ,T_Q }
(i)}{\rm {\bf W}}_{T_Q }(i)=0, \nonumber\\
1 \le q \le (Q - 1).
\end{eqnarray}

To calculate the TX vector, ${\rm {\bf W}}_{T_Q } (i)$, we start with ${\rm
{\bf H}}_{{\rm intf},T_Q } (i)$ which represents all interference channels
from node $T_Q $ to node ($R_1 ,R_2 ,...,R_{(Q - 1)} )$. ${\rm {\bf
H}}_{{\rm intf},T_Q } (i)$ is an $N_A \times (Q - 1)$ matrix, whose
$q^{\rm{th}}$ column is:
\begin{align}
\label{eq2}
{\rm {\bf H}}_{{\rm intf},T_Q }^{{\rm col}} (i,q) =& \left\{ {\sqrt {P_T
G_{R_q ,T_Q } / N_C } {\rm {\bf W}}_{R_q }^H (i){\rm {\bf H}}_{R_q ,T_Q }
(i)} \right\}^H,\notag\\
&1 \le q \le (Q - 1).
\end{align}
Node $T_Q $ then runs a Singular Value Decomposition (SVD) on ${\rm {\bf
H}}_{{\rm intf},T_Q } (i){\rm {\bf H}}_{{\rm intf},T_Q }^H (i)$. Assuming non-increasing order of eigen-values in the SVD result, node $T_Q
$'s TX vector is calculated as:
\begin{eqnarray}
\label{eq3}
&&{\rm {\bf H}}_{{\rm intf},T_Q } (i){\rm {\bf H}}_{{\rm intf},T_Q }^H (i) =
{\rm {\bf U}}_{T_Q } (i){\rm {\bf \Lambda}} _{T_Q } (i){\rm {\bf U}}_{T_Q }^H (i), \\
&&{\rm {\bf W}}_{T_Q } (i) = {\rm {\bf U}}_{T_Q } (i,N_A ).
\end{eqnarray}
Here ${\rm {\bf U}}_{T_Q } (i,N_A )$ is the $(N_A) ^{{\rm th}}$ column in the
matrix ${\rm {\bf U}}_{T_Q } (i)$.

Next we calculate ${\rm {\bf W}}_{R_Q } (i)$ based on both the desired
channel coming from node $T_Q $ and the interference channel coming from
nodes $(T_1 ,T_2 ,...,T_{(Q - 1)} )$. Here the interference channel from
$T_q $ is denoted as $\sqrt {P_T G_{R_Q ,T_q } / N_C } {\rm {\bf H}}_{R_Q
,T_q } (i){\rm {\bf W}}_{T_q } (i)$ with $1 \le q \le (Q - 1)$. The desired
channel from $T_Q $ is $\sqrt {P_T G_{R_Q ,T_Q } / N_C } {\rm {\bf H}}_{R_Q
,T_Q } (i){\rm {\bf W}}_{T_Q } (i)$. Node $R_Q $'s RX vector ${\rm {\bf
W}}_{R_Q } (i)$ can be derived using a MIMO detection algorithm
(zero-forcing is used in the generic SDMA-MAC). Details relating to link contention,
handshaking, and channel information exchange are referred to \cite{1_SPACEMAC, 2_MIMOMAN, 3_NULLHOC}. These
are not considered here as the focus of this work is the achievable
real-world performance of the concurrent SDMA-MAC during the data
transmission phase of the protocol. Admittedly the structure and mechanism
of the contention windows, RTS, CTS, etc., will impact the overall
performance of the network. However, in the interest of maintaining focus we have chosen to defer these issues to a possible follow on contribution.  Therefore care must be taken to incorporate all MAC specific overheads when translating the results to estimate MAC efficiency or throughput performance.

At this juncture it is worth introducing some underlying assumptions or
limitations in our generic SDMA-MAC which also appear in SPACEMAC \cite{1_SPACEMAC, 2_MIMOMAN} and
NullHoc \cite{3_NULLHOC, 4_nullhoc}.
\begin{enumerate}
\item SDMA-MAC often assumes perfect channel estimation in the system \cite{1_SPACEMAC, 2_MIMOMAN}.
\item TX and RX vectors are calculated using the zero-forcing or SVD
based algorithm \cite{1_SPACEMAC, 2_MIMOMAN, 3_NULLHOC, 4_nullhoc}.
\item Each link simply uses a fixed modulation scheme, specifically,
the 2Mbps mode in 802.11b. Multi-rate capabilities embedded within the concurrent links are not fully
utilized.
\item TX vectors are calculated to minimize the resultant interference on the
existing links. But the optimization of SNR within the desired
communication is not considered.
\item Simulations in SPACEMAC and NullHoc consider at most $N_A $ concurrent
links, and the capability of supporting more than $N_A $ links is not
evaluated.
\end{enumerate}

\subsection{Description of the Simulation Environment and Metric Used}

Our simulations are conducted in a single-hop Ad Hoc network, where all
concurrent links are randomly and uniformly placed in a rectangle box of
200m by 200m. Each node is equipped with $N_A=4$ antennas, and uses a
single spatial stream. The system bandwidth, $W$, is assumed to be $W=20$MHz. The modulation is assumed to be OFDM with $N_C=64$
subcarriers and the guard interval is $\rho_G=1/4$. We assume no power
control in the network with the total TX power per node $P_T=25$dBm. Power decay between any two nodes is calculated according to
the simplified path loss model \cite{6_Wireless_Comm_David} with an exponent of 3,
$d_0= 1$m, and wave-length $\lambda= 0.125$m. Fast fading Rayleigh
channels are kept invariant during the transmission period. Background noise power per
subcarrier is $\sigma_N^2=-113$dBm. When link adaptation is
enabled, the link can pick one of the eight modulation and coding schemes
(MCS) shown in Table \ref{Table_I_MCS}. Also, all packets carry the same
amount of data, namely, 100 bytes. We use MATLAB software to build our simulation framework, and each point in our results is an average of 1000 independently generated topologies.

\newcommand{\tabincell}[2]{\begin{tabular}{@{}#1@{}}#2\end{tabular}}

\begin{table}
\caption{List of Modulation and Coding Schemes} \centering \label{Table_I_MCS}
\begin{tabular}{|c||c||c||c|}
\hline \textbf{\tabincell{c}{MCS \\ Index}} & \textbf{QAM Type} & \textbf{\tabincell{c}{Coding\\ Rate}} & \textbf{\tabincell{c}{Minimum required effective \\ PPSNR to achieve target \\ BER/PER (10{\%} PER)}} \\
\hline 0 & \textrm{BPSK} & 1/2 & \textrm{1.4 dB} \\
\hline 1 & \textrm{QPSK} & 1/2 & \textrm{4.4 dB} \\
\hline 2 & \textrm{QPSK} & 3/4 & \textrm{6.5 dB} \\
\hline 3 & \textrm{16QAM} & 1/2 & \textrm{8.6 dB} \\
\hline 4 & \textrm{16QAM} & 3/4 & \textrm{12 dB} \\
\hline 5 & \textrm{64QAM} & 2/3 & \textrm{15.8 dB} \\
\hline 6 & \textrm{64QAM} & 3/4 & \textrm{17.2 dB} \\
\hline 7 & \textrm{64QAM} & 5/6 & \textrm{18.8 dB} \\
\hline
\end{tabular}
\end{table}

We use the network sum-throughput metric \cite{COOP_MAC, 13_Sum_rate_3} in our study which measures
the successfully transmitted throughput summed from all links.
However, in order to decide if a particular link is viable or not, we
calculate the effective Post Processing SNR (PPSNR) \cite{8_PER_model_1} at the receiver for
each link. For a given MCS, if the effective PPSNR is above the minimum
required for the desired QoS (see Table \ref{Table_I_MCS}), then we declare the link viable
and include the link throughput within the sum-throughput calculation.
Otherwise, the link is assumed not to be usable. The effective PPSNR is calculated as follows. Once the TX vector ${\rm {\bf W}}_{T_Q } (i)$ and the RX vector ${\rm {\bf
W}}_{R_Q } (i)$ have been determined for all links in the network, the PPSNR
can be calculated as:
\begin{align}
\label{PPSNR_eq5}
&\Gamma_{R_Q } (i) = \left| {\rm {\bf W}}_{R_Q }^H
(i) \cdot {\rm {\bf H}}_{R_Q ,T_Q }^{{\rm Rec}} (i) \right|^2  \notag\\
&\ \ \ \ \cdot\left\{{\sum\limits_{q = 1,q \ne Q}^K {\left| {\rm
{\bf W}}_{R_Q }^H (i) \cdot {\rm {\bf H}}_{R_Q ,T_q }^{{\rm Rec}} (i) \right|^2} + \sigma _N^2 }\right\}^{-1},\\
\label{H_rec}
&{\rm {\bf H}}_{R_Q ,T_q }^{{\rm Rec}} (i) = \sqrt {P_T G_{R_Q ,T_q } / N_C
} {\rm {\bf H}}_{R_Q ,T_q } (i){\rm {\bf W}}_{T_q } (i).
\end{align}
Using the PPSNR $\mathop \Gamma \nolimits_{R_Q } (i)$ and $\Gamma_{R_Q ,{\rm dB}} (i) = 10\log _{10} \left( {\mathop \Gamma
\nolimits_{R_Q } (i)} \right)$, we then calculate
link $L_Q $'s effective PPSNR $\mathop \Gamma \nolimits_{R_Q ,{\rm dB}}^{\rm eff} $
via Eqn. (\ref{PPSNR_eq6}) \cite{8_PER_model_1} as
\begin{eqnarray}
\label{PPSNR_eq6}
\mathop \Gamma \nolimits_{R_Q ,{\rm dB}}^{\rm eff} = \frac{1}{N_C }\sum\limits^{N_C}_{i = 1} {\Gamma_{R_Q ,{\rm dB}} (i)} - \alpha\cdot var \left[ {\Gamma_{R_Q ,{\rm dB}} (i) } \right].
\end{eqnarray}
Here variance $var$ is calculated over all subcarriers, and $\alpha = 0.125$ is fitted offline \cite{8_PER_model_1}.


\section{Quantifying Performance with Realistic Parameters and Algorithms}

This section is broken down into 4 subsections. In these subsections we separately look at the impact of MMSE, channel estimation error, link
adaptation, and TX beamforming on the sum-throughput of our generic SDMA-MAC. These will be studied in isolation of
one another. Section IV will then evaluate the
SDMA-MAC that incorporates all above four elements.

\subsection{MMSE vs. ZF}

In this section we look at the impact of using the more common MMSE (minimum
mean squared error) detector instead of the idealized ZF (zero-forcing) detector
assumed in the class of SDMA-MACs \cite{1_SPACEMAC, 2_MIMOMAN, 3_NULLHOC, 4_nullhoc}. The reason why the MMSE detector is
more common is because it has the same hardware complexity as the
ZF detector, but does not suffer from the unwanted noise
enhancement properties of the latter \cite{9_ZF_MMSE}. Additionally, one of the drawbacks of the MACs presented in \cite{1_SPACEMAC, 2_MIMOMAN, 3_NULLHOC, 4_nullhoc} is that channel access is sequential with current link not knowing anything about other links that might access the channel after it. In this section we also want to consider the potential benefits of relaxing this assumption. We will refer to this scheme as the Universal-MMSE scheme and describe it in subsection (3).

\subsubsection{Zero-Forcing Detection}

Using the same notation as in section II, link $L_Q $'s RX vector under the
ZF criterion is expressed as:
\begin{align}
\label{ZF_eq8}
&{\rm {\bf W}}_{R_Q } (i) = {\mathbb{N}}\left\{ {{\rm {\bf B}}_{R_Q } (i)\left[
{{\rm {\bf B}}_{R_Q }^H (i){\rm {\bf B}}_{R_Q } (i)} \right]{\rm {\bf e}}_1
} \right\}, \\
&{\rm {\bf e}}_1 = {\underbrace {[1,0,...,0]}_{Q\ {\rm elements}}}^T.
\end{align}
Here ${\rm {\bf B}}_{R_Q } (i)=\left[{\rm {\bf H}}_{R_Q ,T_Q }^{{\rm Rec}} (i), {\rm {\bf H}}_{R_Q ,T_{Q-1} }^{{\rm Rec}} (i), \cdots, {\rm {\bf H}}_{R_Q ,T_1 }^{{\rm Rec}} (i)\right]$ is an $N_A \times Q$ matrix, and ${\rm {\bf H}}_{R_Q ,T_q }^{{\rm Rec}} (i)$ is given in Eqn. (\ref{H_rec}). ${\mathbb{N}}\{ \cdot \}$ denotes the vector normalization with unit power. With the
expression for ${\rm {\bf W}}_{R_Q } (i)$, we then place it in the
equation for PPSNR (Eqn. (\ref{PPSNR_eq5}-\ref{PPSNR_eq6})) to determine if a given link is active or
not. From there we calculate the sum throughput as described in section II.C.


\subsubsection{MMSE Detection}

Link $L_Q $'s RX vector under the MMSE criterion is expressed as:
\begin{align}
\label{MMSE_1}
&{\rm {\bf W}}_{R_Q } (i) = {\mathbb{N}}\left\{ {{\rm {\bf C}}_{R_Q
,{\rm MMSE}}^{ - 1} (i) \cdot {\rm {\bf H}}_{R_Q ,T_Q }^{{\rm Rec}} (i)}
\right\}, \\
\label{MMSE_2}
&{\rm {\bf C}}_{R_Q ,{\rm MMSE}} (i) = \sum\limits_{q = 1}^{Q-1} {{\rm
{\bf H}}_{R_Q ,T_q }^{{\rm Rec}} (i)\left\{ {{\rm {\bf H}}_{R_Q ,T_q
}^{{\rm Rec}} (i)} \right\}^H} + \sigma _N^2 {\rm {\bf I}}_{N_A }.
\end{align}
Here ${\rm {\bf H}}_{R_Q ,T_q }^{{\rm Rec}} (i)$ is given in Eqn. (\ref{H_rec}). Similar to ZF receiver, the derived ${\rm {\bf W}}_{R_Q }
(i)$ is used in the PPSNR calculation (Eqn. (\ref{PPSNR_eq5}-\ref{PPSNR_eq6})) and subsequently in the
sum-throughput calculation.

\subsubsection{Universal MMSE}

Previous derivations for link $L_Q $'s RX vector only consider the
interference channel coming from link $L_1 $ to $L_{(Q - 1)} $. The
residual interference caused by $L_{(Q + 1)}$ to $L_K$ is not
considered. This can  have a negative impact on the PPSNR results. Here we look to answer the question of how much the performance of the system might be improved if the receive beamnulling was performed at each node with full knowledge of all $K$ transmitters. Firstly, link $L_Q$ estimates the covariance of the
received signal from all $K$ TX nodes as:
\begin{align}
\label{MMSE_3}
&{\rm {\bf C}}_{R_Q ,{\rm UMMSE}} (i) =\notag\\
&\sum\limits_{q = 1, q\neq Q}^K {{\rm {\bf
H}}_{R_Q ,T_q }^{{\rm Rec}} (i)\left\{ {{\rm {\bf H}}_{R_Q ,T_q
}^{{\rm Rec}} (i)} \right\}^H} + \sigma _N^2 {\rm {\bf I}}_{N_A } .
\end{align}
Based on the estimate ${\rm {\bf C}}_{R_Q ,{\rm UMMSE}} (i)$, the
corresponding Universal MMSE RX vector is:
\begin{equation}
\label{MMSE_4}
{\rm {\bf W}}_{R_Q } (i) = {\mathbb{N}}\left\{ {{\rm {\bf C}}_{R_Q
,{\rm UMMSE}}^{-1}(i)\cdot{\rm {\bf
H}}_{R_Q ,T_Q }^{{\rm Rec}} (i)} \right\}.
\end{equation}
Again, ${\rm {\bf W}}_{R_Q }(i)$ is used to calculate the PPSNR (Eqn. (\ref{PPSNR_eq5}-\ref{PPSNR_eq6})) and the
sum-throughput.


\subsubsection{Simulation Results}

\begin{figure}
\centering
\includegraphics[width=3.3in]{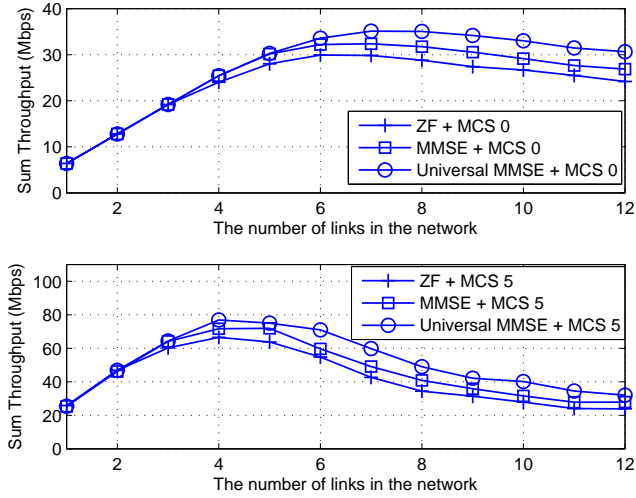}
\caption{SDMA-MAC's throughput performance comparing a ZF, an MMSE, and a Universal MMSE RX vector calculation algorithm.}
\label{Fig1_Rx_vector}
\end{figure}

We now simulate the SDMA-MAC protocol using both the ZF and the MMSE
algorithms for the RX vector ${\rm {\bf W}}_{R_Q } (i)$. We assume perfect
channel estimation, and as in the case of \cite{1_SPACEMAC, 2_MIMOMAN, 3_NULLHOC} we fix the MCS for each link
to either MCS 0 or MCS 5. Fig. \ref{Fig1_Rx_vector} shows the sum-throughput as a function of
the number of concurrent links allowed by the MAC. As expected, initially
the sum-throughput increases with the number of concurrent links in the
network, however, as additional concurrent links are added the interference
power dominates, thus causing a decrease in the network sum-throughput.
Fig. \ref{Fig1_Rx_vector} also shows that the MMSE receiver outperforms the ZF receiver by an average of 10{\%}, and a maximum of 20{\%}. The Universal MMSE scheme outperforms the ZF receiver by up to 40{\%}. This is because for link $L_Q$, the Universal MMSE protocol takes into account the residual interference from all links irrespective of the order in which they start transmission, whereas the ZF and MMSE solutions only take into account the subset of links that accessed the channel before link $L_Q$.

\subsection{Impact of Channel Estimation Errors}

Channel estimation errors impact the sum throughput by increasing interference and also reducing the PPSNR. Given that the SDMA-MAC uses MIMO beamforming at both the TX and the RX, estimation errors will impact both the TX vector ${\rm {\bf W}}_{T_Q } (i)$
and the RX vector ${\rm {\bf W}}_{R_Q } (i)$. Let us first derive the expression of the noisy RX vector ${\rm {\bf
W}}_{R_Q } (i)$. Note that this is calculated using either the ZF method
(Eqn. (\ref{ZF_eq8})) or the MMSE method (Eqn. (\ref{MMSE_1}-\ref{MMSE_4})). These calculations all rely
on the following channel information ${\rm {\bf H}}_{R_Q ,T_q }^{{\rm Rec}}
(i)$. Under imperfect channel estimation, the noisy estimate of ${\rm {\bf
H}}_{R_Q ,T_q }^{{\rm Rec}} (i)$ is given by:
\begin{align}
\label{eq14}
\widetilde{{\rm {\bf H}}}_{R_Q ,T_q }^{{\rm Rec}} (i) =& \sqrt {P_T G_{R_Q
,T_q } / N_C } {\rm {\bf H}}_{R_Q ,T_q } (i){\rm {\bf W}}_{T_q } (i)\nonumber\\
&+ \sqrt{\sigma _C^2 } {\rm {\bf Z}}_{R_Q ,T_q } (i), 1 \le q \le K.
\end{align}
Here ${\rm {\bf Z}}_{R_Q ,T_q } (i)$ represents the channel estimation noise
which is modeled as a complex Gaussian random variable with zero mean and
unit variance. $\sigma _C^2 $ is the variance of the estimation noise, which
is dependent on the variance of the background noise per subcarrier $\sigma
_N^2 $ (e.g., under $L$ training symbols and least-square estimation,
$\sigma _C^2 $ is equal to $\sigma _N^2 / L$ \cite{3_NULLHOC}). In this way, the noisy RX
vector ${\rm {\bf W}}_{R_Q } (i)$ is derived by using noisy estimate
$\widetilde{{\rm {\bf H}}}_{R_Q ,T_q }^{{\rm Rec}} (i)$ rather than perfect
estimate ${\rm {\bf H}}_{R_Q ,T_q }^{{\rm Rec}} (i)$ in ZF or MMSE methods.



Now we look at the derivation of the noisy TX vector ${\rm {\bf W}}_{T_Q }
(i)$. Since ${\rm {\bf W}}_{T_Q } (i)$ depends on ${\rm {\bf W}}_{R_q } (i)$
($q=1, \ldots, Q-1$), then the noisy estimate of ${\rm {\bf W}}_{T_Q } (i)$
will be a function of the noisy estimates of other ${\rm {\bf W}}_{R_q }
(i)$. Recalling Eqn. (\ref{eq3}) ${\rm {\bf W}}_{T_Q } (i) = {\rm {\bf U}}_{T_Q }
(i,N_A )$, we have that ${\rm {\bf W}}_{T_Q } (i)$ is a function of the
SVD results of ${\rm {\bf H}}_{{\rm intf},T_Q } (i){\rm
{\bf H}}_{{\rm intf},T_Q }^H (i)$. In the presence of channel estimation
errors, ${\rm {\bf H}}_{{\rm intf},T_Q } (i)$'s column, ${\rm {\bf
H}}_{{\rm intf},T_Q }^{{\rm col}} (i,q)$ in Eqn. (\ref{eq2}), will be replaced by
the noisy estimate:
\begin{align}
\label{eq15}
\widetilde{{\rm {\bf H}}}_{{\rm intf},T_Q }^{{\rm col}} (i,q) =& \left\{
{\sqrt {P_T G_{R_q ,T_Q } / N_C } {\rm {\bf W}}_{R_q }^H (i){\rm {\bf
H}}_{R_q ,T_Q } (i)} \right\}^H \nonumber\\
&+ \sqrt {\sigma _C^2 } {\rm {\bf Z}}_{T_Q,R_q } (i).
\end{align}
Again, here ${\rm {\bf Z}}_{T_Q ,R_q } (i)$ represents the channel
estimation noise. Using the noisy estimate $\widetilde{{\rm {\bf
H}}}_{{\rm intf},T_Q }^{{\rm col}} (i,q)$, the noisy TX vector ${\rm {\bf
W}}_{T_Q } (i)$ is calculated according to Eqn. (\ref{eq3}).

Finally, given the resulting noisy TX vectors and RX vectors, the effective
PPSNR can be derived using Eqn. (\ref{PPSNR_eq5}-\ref{PPSNR_eq6}), and the sum-throughput performance
can be evaluated accordingly.

\subsubsection{Simulation Results}

\begin{figure}
\centering
\includegraphics[width=3.3in]{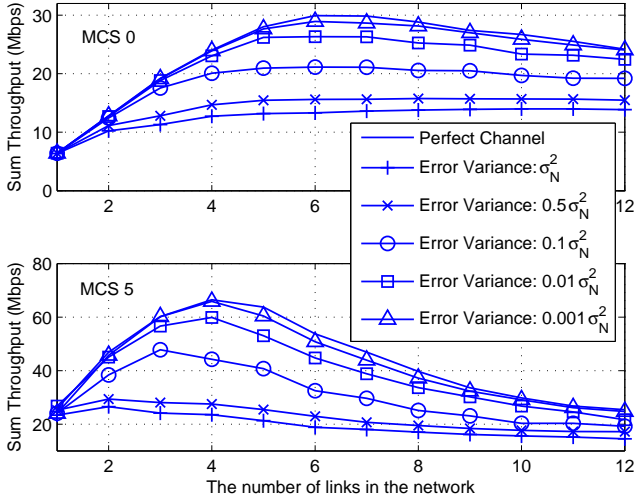}
\caption{Sum throughput performance of SDMA-MAC under different channel estimation errors. Assume that each link uses either MCS 0 or MCS 5.}
\label{Fig2_estimation_error}
\end{figure}

We simulate the sum throughput of the SDMA-MAC protocol under different
channel estimation errors. Here the variance of the estimation error is set
to $\sigma _N^2 $, $0.5\sigma _N^2 $, $0.1\sigma _N^2 $, $0.01\sigma _N^2 $
and $0.001\sigma _N^2 $, respectively, where $\sigma _N^2 $ denotes the
power of the background noise per subcarrier and is equal to -113dBm in our study. Each
link's MCS is fixed as MCS 0 or MCS 5, and simulation results under
different number of concurrent links are plotted in Fig. \ref{Fig2_estimation_error}. The curves in
the figure show that, compared with the result under perfect channel
estimation, system's sum throughput is seriously degraded when the
estimation variance is $\sigma _N^2 $ or $0.5\sigma _N^2 $. Meanwhile, even
under estimation variance of $0.1\sigma _N^2 $ and $0.01\sigma _N^2 $, there
still exists considerable performance loss in the sum throughput. Generally, it is safe to
assume that for any Ad Hoc system, the estimation noise variance will be at
best 0.1$\sigma_{N}^{2}$.

\subsection{Impact of Link Adaptation}

\subsubsection{Link Adaptation Design}

There are 8 different MCSes in this paper, and each
link can adaptively select the proper MCS based on the estimated PPSNR $\mathop {\widehat{\Gamma }}\nolimits_{R_Q } (i)$. The derivation of $\mathop {\widehat{\Gamma }}\nolimits_{R_Q } (i)$ is based on Eqn. (\ref{PPSNR_eq5}) but using the estimated channel information as shown in Eqn. (\ref{eq14}). To
study the link adaptation in isolation, this subsection assumes perfect channel
estimates ($\sigma _C^2 = 0$). Given ${\widehat{\Gamma }}_{R_Q ,{\rm dB}} (i) = 10\log _{10} \left(
{\mathop {\widehat{\Gamma }}\nolimits_{R_Q } (i)} \right)$, link $L_Q $'s effective SNR is estimated as:
\begin{align}
\label{eq16}
 {\widehat{\Gamma }}_{R_Q,{\rm dB}}^{\rm eff} = & \frac{1}{N_C}\sum\limits_{i=1}^{N_C}{\widehat{\Gamma }}_{R_Q ,{\rm dB}} (i) - \alpha \cdot var\left[{\widehat{\Gamma }}_{R_Q ,{\rm dB}} (i)\right] \nonumber\\
 &- \Gamma _{L_Q }^{\rm Backoff},\ \ \ \ \alpha = 0.125
\end{align}
Later, link $L_Q $ will select the
highest MCS whose threshold listed in Table I is smaller than the estimated SNR $\mathop {\widehat{\Gamma }}\nolimits_{R_Q
,{\rm dB}}^{\rm eff} $. Finally, $\Gamma _{L_Q }^{\rm Backoff} $ in Eqn. (\ref{eq16}) is a correction term that makes up for
the inaccuracy of the PPSNR estimation (e.g., due to imperfect channel estimation). Its value can be
tuned at run-time using the real packet error rate embedded within the
ACK packet.


\subsubsection{Simulation Results}

\begin{figure}
\centering
\includegraphics[width=3.3in]{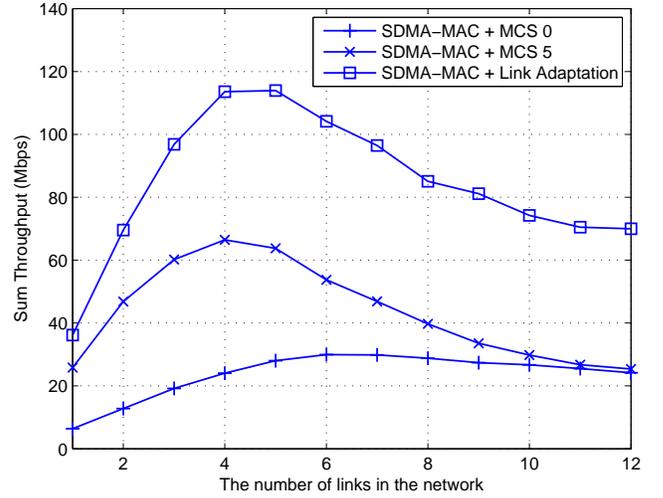}
\caption{Sum throughput performance of SDMA-MAC under the usage of link adaptation. Assume perfect channel estimation and ZF based RX vector derivation.}
\label{Fig3_link_adaptation}
\end{figure}

We evaluate the sum throughput performance of our generic SDMA-MAC by using
the link adaptation process discussed above. The results are summarized in
Fig. \ref{Fig3_link_adaptation}. Here we assume perfect channel estimation in the system, and RX
vectors are derived via the ZF method. For completeness, we also provide the
results of fixed MCS selection (MCS 0 or MCS 5) in that figure. The
resultant curves underscore the importance of link adaptation in improving
the network performance. Compared with the fixed MCS selection (MCS 0 or MCS
5) under total 4 concurrent links, the usage of link adaptation can provide
additional throughput gains of around 70{\%} (for MCS 5) to 200{\%} (for MCS
0).

\subsection{Combining TX Beamforming with SDMA-MAC}

In the generic SDMA-MAC, the TX node $T_Q $ of link $L_Q $ knows nothing
about the channel between TX node $T_Q $ and RX node $R_Q $. In this
subsection we pose the question of how the performance could be improved if
node $T_Q $ knew about the channel between $T_Q $ and $R_Q $, and was able
to beamform accordingly. Consider the derivation of ${\rm {\bf W}}_{T_Q } (i)$ in Eqn. (\ref{eq3}),
provided that $Q \le N_A $, there will be $(N_A - Q + 1)$ candidates for the
TX vector ${\rm {\bf W}}_{T_Q } (i)$ (${\rm {\bf U}}_{T_Q } (i,Q),{\rm {\bf
U}}_{T_Q } (i,Q + 1),...,{\rm {\bf U}}_{T_Q } (i,N_A ))$, which can all
satisfy the orthogonality condition of Eqn. (\ref{eq1}). Besides, any linear
combination of these candidates is also orthogonal with existing links (Eqn.
(\ref{eq1})). This observation indicates that we can choose an optimized linear
combination of these candidates, so that the resultant PPSNR in the desired
communication is improved. We name this scheme TX beamforming to distinguish
it from the TX beamnulling scheme (Eqn. (\ref{eq3})) used in the baseline SDMA-MAC. We apply TX
beamforming only to links $L_Q $ with $Q \le N_A $, all other links (link
$L_{N_A + 1} $ to link $L_K )$ will use the default TX beamnulling of the
SDMA-MAC.

\subsubsection{TX Beamforming Calculation}

Consider link $L_Q $ with $Q \le N_A $, we use ${\rm {\bf U}}_{T_Q
}^{{\rm INIT}} (i)$ to denote all the TX vector candidates at node $T_Q $.
This is an $N_A \times (N_A - Q + 1)$ matrix composed of columns ${\rm {\bf
U}}_{T_Q } (i,Q),{\rm {\bf U}}_{T_Q } (i,Q + 1),...,{\rm {\bf U}}_{T_Q }
(i,N_A )$. The resultant TX vector is given as ${\rm {\bf W}}_{T_Q } (i) =
{\rm {\bf U}}_{T_Q }^{{\rm INIT}} (i){\rm {\bf D}}_{T_Q } (i)$, where ${\rm
{\bf D}}_{T_Q } (i)$ is an $(N_A - Q + 1)\times 1$ column vector with ${\rm
{\bf D}}_{T_Q }^H (i){\rm {\bf D}}_{T_Q } (i) = 1$ representing the linear
combination of ${\rm {\bf U}}_{T_Q }^{{\rm INIT}} (i)$. Given a specific
${\rm {\bf D}}_{T_Q } (i)$, the calculated PPSNR at the $i$th subcarrier of
link $L_Q $ under the MMSE criterion is given as:
\begin{align}
\label{eq18}
\Gamma _{R_Q } (i) =& (P_T G_{R_Q ,T_Q } /N_C)\left\{ {{\rm {\bf
H}}_{R_Q ,T_Q } (i){\rm {\bf U}}_{T_Q }^{{\rm INIT}} (i){\rm {\bf D}}_{T_Q
} (i)} \right\}^H \notag\\
&{\rm {\bf C}}_{R_Q ,{\rm MMSE}}^{ - 1} (i){\rm {\bf H}}_{R_Q ,T_Q
} (i){\rm {\bf U}}_{T_Q }^{{\rm INIT}} (i){\rm {\bf D}}_{T_Q } (i).
\end{align}
Here ${\rm {\bf C}}_{R_Q,{\rm MMSE}}^{ - 1} (i)$ is given in Eqn. (\ref{MMSE_2}). Obviously,
the optimal linear combination vector ${\rm {\bf D}}_{T_Q } (i)$ can be
calculated as the maximum eigen-vector of $\left\{ {{\rm {\bf H}}_{R_Q ,T_Q
} (i){\rm {\bf U}}_{T_Q }^{{\rm INIT}} (i)} \right\}^H{\rm {\bf C}}_{R_Q
,{\rm MMSE}}^{ - 1} (i){\rm {\bf H}}_{R_Q ,T_Q } (i){\rm {\bf U}}_{T_Q
}^{{\rm INIT}} (i)$ that is corresponding to the maximum eigen-value. And
the associated TX vector ${\rm {\bf W}}_{T_Q } (i)$ can be calculated
according to the optimal ${\rm {\bf D}}_{T_Q } (i)$ as ${\rm {\bf W}}_{T_Q } (i) =
{\rm {\bf U}}_{T_Q }^{{\rm INIT}} (i){\rm {\bf D}}_{T_Q } (i)$.

\subsubsection{Simulation Results}

\begin{figure}
\centering
\includegraphics[width=3.3in]{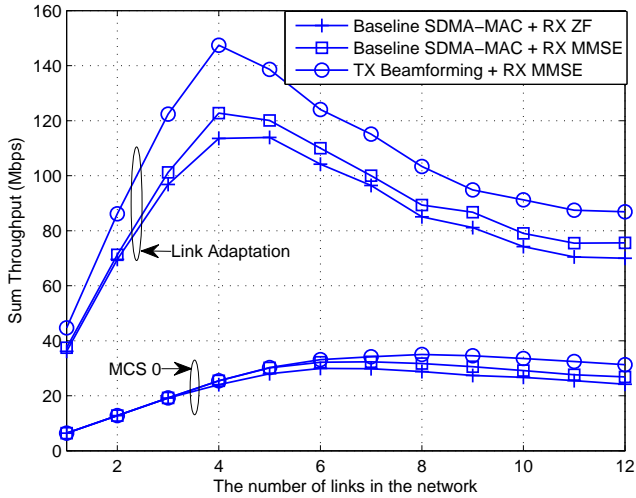}
\caption{Sum throughput performance under our discussed TX beamforming design. Assume perfect channel estimation in the network, and each link's MCS is either fixed as MCS 0, or adaptively tuned.}
\label{Fig4_Tx_beamforming}
\end{figure}

We evaluate the sum throuhgput performance of our reference SDMA-MAC with
the inclusion of the TX beamforming approach combined with MMSE RX vectors. By way of comparison we
also provide the throughput results of two other schemes, the first includes
baseline SDMA-MAC with ZF RX vectors (Eqn. (\ref{ZF_eq8})), and the second includes baseline SDMA-MAC with MMSE RX vectors (Eqn. (\ref{MMSE_1}-\ref{MMSE_2})). We assume perfect channel estimation in the network,
and the MCS in each link is either fixed at MCS 0, or varied under the
link adaptation protocol. The sum throughput results are
shown in Fig. \ref{Fig4_Tx_beamforming}. Compared with the performance of the baseline SDMA-MAC
with ZF RX vectors, the introduction of TX
beamforming can have around 20{\%} improvement in terms of sum throughput.
When compared with baseline SDMA-MAC plus MMSE RX vectors,
the TX beamforming design can still have more than 10{\%} throughput gain.

\section{Combined Performance Characterization}

After evaluating the impact of each of the four elements introduced in this
paper in isolation, we now look to compare the performance of the baseline
concurrent SDMA-MAC with the variant that includes the following four
elements: (a) practical MMSE algorithm; (b) channel estimation errors; (c) link
adaptation mechanism; (d) incorporation of TX beamforming. For comparison
purposes, this section also introduces results of a non-concurrent MAC (only one link is allowed at any given time) that can employ any number of spatial streams less than $N_A$. The non-concurrent MAC also employs  MIMO TX and RX beamforming, link adaptation and channel estimation
errors. Detailed settings in these MAC schemes are summarized in Table \ref{Table_II_MAC}.

\begin{table*}
\caption{Detailed Settings in the Considered MAC Schemes} \centering \label{Table_II_MAC}
\begin{tabular}{|c|c|c|c|c|c|}
\hline \textbf{MAC Scheme} & \textbf{TX Vectors} & \textbf{RX Vectors} & \textbf{\tabincell{c}{Link Adaptation \\ (\# of streams)}} & \textbf{\tabincell{c}{Link Adaptation \\ (MCS per stream)}} & \textbf{\tabincell{c}{Channel \\ Estimation \\ Error}} \\

\hline \textbf{\tabincell{c}{Baseline Concurrent\\ SDMA-MAC}} & \tabincell{c}{TX Beamnulling \\ Eqn. (\ref{eq3})} & \tabincell{c}{ZF method \\ Eqn. (\ref{ZF_eq8})} & \tabincell{c}{Each link uses \\ only one stream.} & \tabincell{c}{The MCS per \\ link is fixed \\ as MCS 0.} & {0.1$\sigma_N^2$} \\

\hline \textbf{\tabincell{c}{Enhanced (realistic) \\ SDMA-MAC Scheme}} & \tabincell{c}{TX Beamforming \\ Section III.D} & \tabincell{c}{MMSE method\\ Eqn. (\ref{MMSE_1}-\ref{MMSE_2})} & \tabincell{c}{Each link uses \\ only one stream.} & \tabincell{c}{The MCS per link \\ is adaptively selected. \\ (section III.C)} & {0.1$\sigma_N^2$} \\

\hline \textbf{\tabincell{c}{Enhanced Scheme \\ with the Universal \\ MMSE Scheme \\ at the RX}} & \tabincell{c}{TX Beamforming \\ Section III.D} & \tabincell{c}{Universal \\ MMSE method \\ Eqn. (\ref{MMSE_3}-\ref{MMSE_4})} & \tabincell{c}{Each link uses \\ only one stream.} & \tabincell{c}{The MCS per link \\ is adaptively selected. \\ (section III.C)} & {0.1$\sigma_N^2$} \\

\hline \textbf{\tabincell{c}{Non-concurrent \\ MAC Scheme}} & \tabincell{c}{SVD based \\ method \\ see \cite{14_SVD}} & \tabincell{c}{SVD based \\ method \\ see \cite{14_SVD}} & \tabincell{c}{Each link can use \\ up to $N_A$  streams. \\ The number of \\ streams is adaptively \\ selected to maximize \\ the throughput.} & \tabincell{c}{The MCS per stream \\ is adaptively selected, \\ which is similar to \\ the method in \\ section III.C.} & {0.1$\sigma_N^2$} \\

\hline
\end{tabular}
\end{table*}

\subsection{Simulation Results}

\begin{figure}
\centering
\includegraphics[width=3.3in]{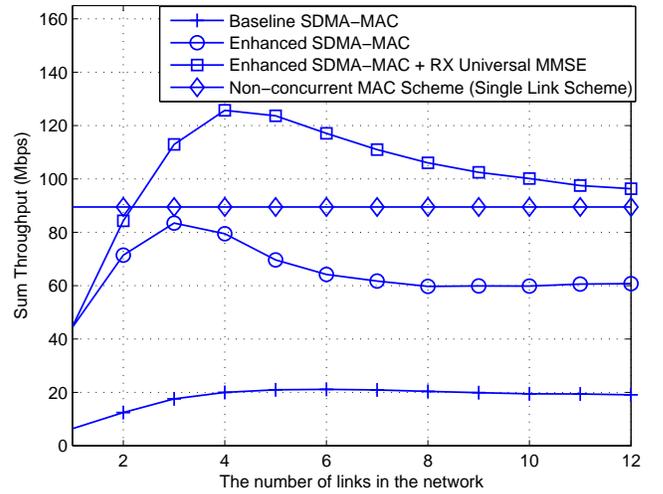}
\caption{Sum throughput performance under different MAC schemes. Channel estimation error is fixed as 0.1$\sigma_N^2$.}
\label{Fig5_MAC_scheme}
\end{figure}

The simulation results are summarized in Fig. \ref{Fig5_MAC_scheme}. Firstly, the baseline SDMA-MAC has the lowest sum throughput, which is mostly
due to the lack of link adaptation in it. Secondly, the enhanced SDMA-MAC has 3x to 4x higher throughput than the baseline SDMA-MAC, but its results are lower than that of non-concurrent MAC. It is because that under imperfect channel estimation, residual interference among concurrent links has a significant impact on the overall network performance. This is partly the reason why  the enhanced design with the Universal MMSE scheme has the highest sum throughput. With 4 concurrent links it shows a 500{\%} improvement over the baseline SDMA-MAC and 40{\%} improvement over the non-concurrent MAC scheme. The superior performance of our enhanced SDMA MAC over the non-concurrent MAC scheme is primarily attributed to the use of multiple antennas for spatial interference mitigation and the associated power allocation.

At this juncture it is worth noting that in an actual fully functioning MAC, the overhead of the contention window is most likely highest for the enhanced scheme with the Universal MMSE, and is smallest for the non-concurrent MAC. This is because the Universal-MMSE method will require more control packets in order to get all the information needed.


\section{Conclusion}

The aim of this work was to investigate how an SDMA MAC based on the notion of concurrent links would perform in a real operating network that is subject to self interference and channel estimation errors that will negatively impact the performance. We also looked to bring in link adaptation and MMSE based beamforming that are part of almost any operating MIMO based system. Our work uncovered two rather important results. The first, is that significant performance improvements can be had with the combination of MMSE based beamforming and link adaptation, even in the presence of channel estimation errors. The second
is that a single link transmission strategy that can use multiple spatial streams is rather hard to beat with a concurrent transmission strategy that looks to maximize the number of transmissions each with a single spatial stream.



%





\ifCLASSOPTIONcaptionsoff
  \newpage
\fi



%



\bibliographystyle{IEEEtran}
\bibliography{IEEEabrv,tvt_draft_final}

\end{document}